# ANOTHER FLATTENED DARK HALO:
# POLAR RING GALAXY A0136−0801

Penny D. Sackett
Institute for Advanced Study, Princeton, NJ 08540
psackett@guinness.ias.edu

Richard W. Pogge
Ohio State University, Columbus, Ohio 43210-1106
pogge@payne.mps.ohio-state.edu

## ABSTRACT

Knowledge of the shape of dark matter halos is critical to our understanding of galaxy formation, dynamics, and of the nature of dark matter itself. Polar ring galaxies (PRGs) — early-type galaxies defined by their outer rings of gas, dust and stars on orbits nearly perpendicular to those of the central host — provide a rare probe of the vertical-to-radial axis ratio ($q_\rho = c/a$) of dark halos. We present a Fabry-Perot velocity field for the H$\alpha$ gas in the kinematically-confirmed PRG A0136−0801. By comparing ring orbits evolved in a generalized mass model to the observed ring velocity field and morphology of A0136−0801, we conclude that $q_\rho \sim 0.5$ and rule out a spherical geometry.

### 1. WHY POLAR RINGS?

Many PRG, including A0136−0801, appear to be at or near equilibrium, so that the ring gas can be assumed to follow closed orbits. The shape and velocity structure of closed orbits contain information about the shape of the underlying mass distribution: orbits are elongated along the short axis of a flattened potential and have speeds that vary with ring azimuth (slower over the poles). Given a viewing geometry, the unique relationship between orbit position and line-of-sight velocity allows one to use the 2-D velocity field of a polar ring to constrain the shape of its dark halo.

A0136−0801 is particularly suitable since its ring rotation curve is flat to more than 12 disk scale lengths above the plane of its galactic disk (Schweizer *et al.* 1983, SWR), indicating that the ring dynamics are dominated by the dark halo. We can therefore expect that the flattening we infer from our Fabry-Perot data reflects the flattening of the *dark* mass. If the mass distribution is not spherical and the polar axis does not lie in the sky plane, the kinematic major and minor axes of the ring will be twisted relative to the morphological axes (Sackett 1991). This signature is in our data (Fig. 1), and has the opposite sense as would be expected for deviations in the velocity field due to inflow. Our goal is to determine the value of the axis ratio of the density distribution, $q_\rho$, that produces the best fit to the geometry and velocity field of the ring.

### 2. THE TWO-DIMENSIONAL VELOCITY FIELD

We used the Rutgers CCD Fabry-Perot spectrograph on the CTIO 4m

telescope to obtain a velocity cube of A0136−0801 in the Hα emission line. Seventeen 900-sec interferograms, sampled in 1Å intervals, were taken with the ∼2.3Å resolution (FWHM) etalon. Median seeing over the ∼4 hour run was ∼1.″5; atmospheric transparency was excellent. The processed interferograms were stacked into a data cube for calibration and analysis using the OASIS Fabry-Perot reduction package developed at OSU (Pogge 1991).

Since the polar ring is completely filled with HII regions, we are able to derive the (relative) Hα emission-line flux and radial velocity for each of the over 2000 pixels in the ring (Fig. 1). The measured Hα emission-line centroids are subject to two principal sources of uncertainty: signal-to-noise (S/N) in each spectrum pixel and an etalon drift that introduces an additional random error of about $\pm 7$ km sec$^{-1}$. Adding the errors in quadrature yields a total median uncertainty of 12 km s$^{-1}$ in the heliocentric velocities.

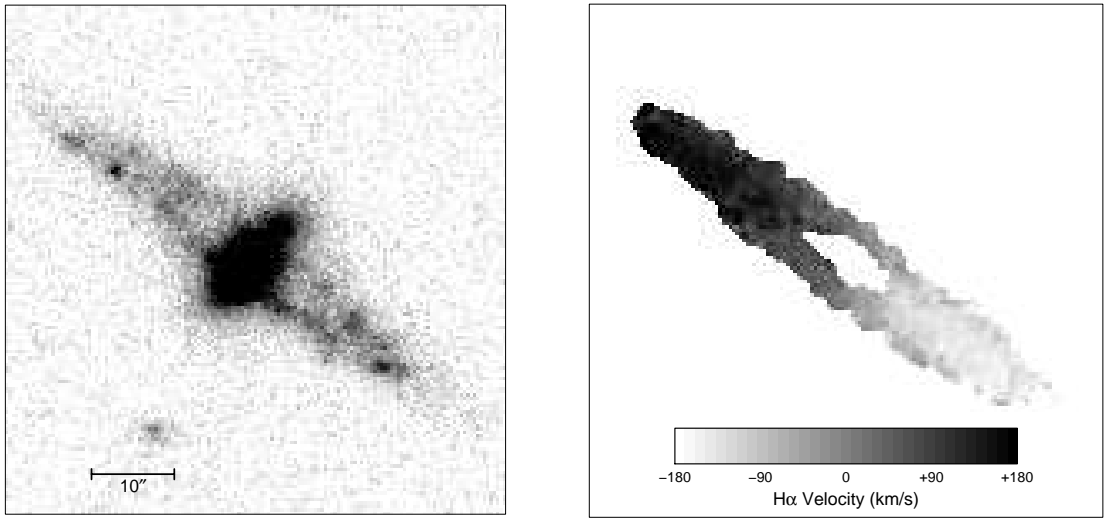

Fig. 1. *Left:* Narrow band image of A0136−0801 created by integrating the Hα data cube along the wavelength axis; image scale is ∼ 0″.4 pixel$^{-1}$. *Right:* Ring velocity field; greyscale indicates the relative velocity scale.

## 3. THE MASS MODEL

The ring gas is assumed to follow closed, co-planar orbits in the gravitational potential of an axisymmetric, oblate, flattened mass distribution, with radial profile of the standard pseudo-isothermal form,

$$\rho(R, z) = \frac{\rho_o \, r_c^2}{r_c^2 + R^2 + (z^2/q_\rho^2)} \quad ,$$

and isodensity surfaces that are stratified on concentric, similar ellipsoids of axis ratio $q_\rho = c/a$. Such a mass distribution produces asymptotically flat rotation curves (with $v_\infty^2 = 4\pi G \rho_o \, r_c^2 \, (1 - q_\rho^2)^{-1/2} q \arccos q$) in the midplane of the potential, which is assumed here to be perpendicular to the polar axis.

We use a shooting algorithm to compute discrete, closed polar orbits in this mass model. After projection onto the sky plane, a cloud-in-cell algorithm is used to create a smooth velocity field with the resolution of the data. This model velocity field is then compared to the data via the $\chi^2$ statistic; each pixel is weighted by the uncertainty in its measured velocity. The free parameters of a model are varied to minimize $\chi^2$ and thus achieve the best *kinematic* fit to the data. With $q_\rho$ held fixed at values in the range $0.2 \leq q_\rho \leq 1$ (corresponding to the range E8 (flat) to E0 (round) in the notation of elliptical galaxies), the free parameters are: $r_c$, $v_\infty$, the systemic velocity $v_{\text{sys}}$, and the projection angles, $\alpha$, $\beta$, and $\gamma$, which control the pivot of the ring about its long axis, the tip of the long ring axis out of the sky plane, and the position angle of the ring on the sky.

For each halo flattening, the fitting algorithm chooses the viewing geometry that best reproduces the kinematic data, without imposing constraints from the ring shape. Only after fitting the kinematics do we examine the *morphological* consequences of varying $q_\rho$.

## 4. EVIDENCE FOR A FLATTENED HALO

The *kinematic* data alone favor models with density axis ratios of $0.5 \leq q_\rho \leq 0.9$ (Fig. 2), regardless of whether the data are weighted equally or by their statistical uncertainties, and regardless of whether or not minor axis data are masked. Only the flatter models with $q_\rho \leq 0.5$, however, are able to reproduce the observed *morphology* of the ring (Figs. 2 & 3). Spherical models provide extremely poor fits to both the kinematics and geometry of the ring. We conclude that it is likely that the mass distribution probed by the polar ring of A0136−0801 is flattened with an axis ratio $q_\rho \sim 0.5$. This result is marginally consistent with previous work for this galaxy based on long-slit data (SWR, Whitmore *et al.* 1987), and with the flattened halos derived for many other systems (for reviews, see Sackett *et al.* 1994 and Rix 1995).

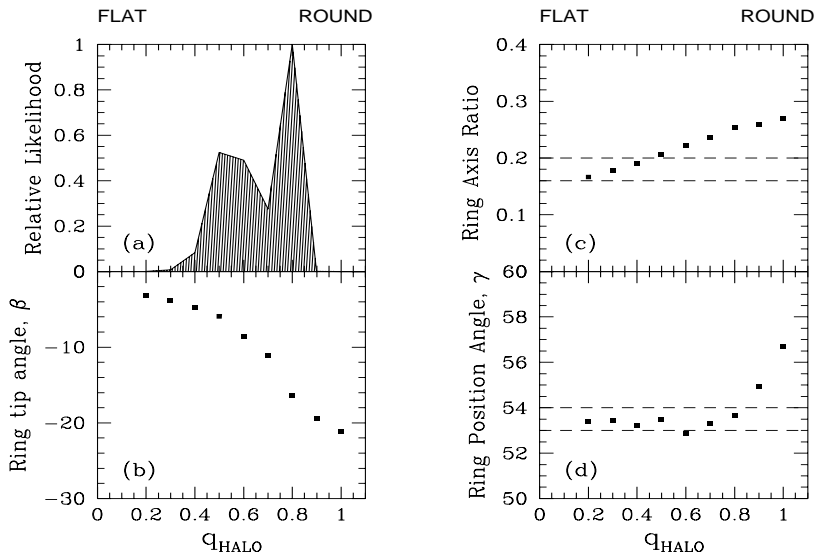

Fig. 2. The results of a typical run. Shown as a function of halo flattening: (a) relative likelihood of the *kinematic* fit, (b) ring tip angle, $\beta$, (c) resulting axis ratio of ring on sky, and (d) ring position angle. The dashed lines in (c) and (d) indicate the range of observed values.

An independent measurement of $\beta$, the angle between the polar axis and the sky, would further constrain $q_\rho$ (Fig. 2). The visual photometry of SWR and the K-band photometry of Arnaboldi *et al.* (1994) may indicate the presence of an edge-on disk embedded in a more spheroidal central component of A0136−0801. If confirmed, this would implicate a small ring tip angle, a flatter halo, and an intrinsic flattening of the spheroidal component similar to that of the dark halo, a situation already noted for NGC 4650A (Sackett *et al.* 1994).

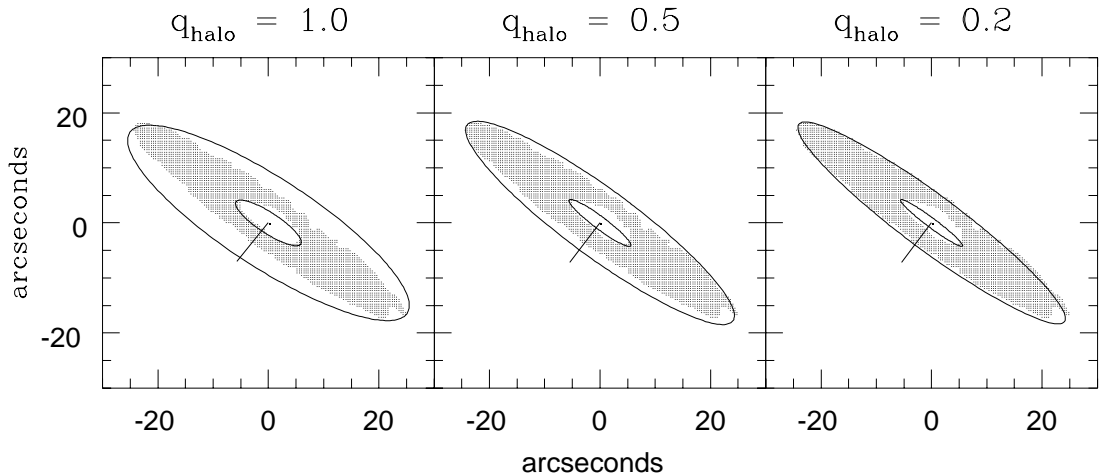

Fig. 3. Projected model orbits (solid lines) at 7″ and 30″ superposed on the masked ring data for the run shown in Fig. 2.

We will present elsewhere (Sackett & Pogge 1995, in preparation) a more complete analysis that examines the effects of warping and different choices for the parameterization of the gravitational potential.

## ACKNOWLEDGEMENTS

Work by PDS was supported by the J. Seward Johnson Charitable Trust and NSF grant AST 92-15485. CTIO observing was supported in part by AURA and travel to Chile by OSU start-up funds. Support for RWP was also provided by NSF grants AST 88-22009 and AST 91-12879.